\documentclass[aps,prl,twocolumn]{revtex4}

\usepackage{times}
\usepackage{amsmath}
\usepackage{psfrag}
\usepackage{epsfig}

\newcommand{\nn}{\nonumber \\}
\newcommand{\ov}[1]{\overline{#1}}

\newcommand{\eqsand}[2]{Eqs.~(\ref{#1}) and (\ref{#2})}

\newcommand{\real}{\mathrm{Re}\,}
\newcommand{\gev}{\,\mbox{GeV}}

\newcommand{\BDTN}{B \to D \tau \nu_{\tau}}

\newcommand{\BDLN}{B \to D \ell \nu_\ell}

\newcommand{\BTN}{B \to \tau \nu_\tau}
\newcommand{\BTNp}{B^+ \to \tau^+ \nu_\tau}

\begin{document}
\title{{\small TTP08-06\hfill SFB/CPP-08-11}\\[1cm]
\boldmath Charged--Higgs effects in a new $B\to D \tau \nu_{\tau}$ differential decay distribution \unboldmath}

\author{Ulrich Nierste} 
\author{St\'ephanie Trine} 
\author{Susanne Westhoff} 

\affiliation{
~\\ 
Institut f\"ur Theoretische Teilchenphysik\\
Karlsruhe Institute of Technology, 76128 Karlsruhe, Germany
}

\begin{abstract}
We show that the decay mode $\BDTN$ is competitive with and complementary
to $\BTN$ in the search for charged--Higgs effects.    
Updating the relevant form factors, we find that the differential distribution in the
 decay chain $\bar{B}\to D\bar{\nu}_{\tau}\tau^-[\to\pi^-\nu_{\tau}]$
 excellently discriminates between Standard--Model and charged--Higgs
 contributions.  By measuring the $D$ and $\pi^-$ energies and the angle
 between the $D$ and $\pi^-$ three-momenta one can determine 
 the effective charged--Higgs coupling including a 
possible CP--violating phase.
\end{abstract}

\maketitle

\section{Introduction}    
The $B$ factories BABAR and BELLE have
accumulated enough statistics to probe extensions of the Higgs sector of
the Standard Model. Notably, the decay $\BTNp$ allows us to place useful
constraints on the parameters $\tan \beta$ and $M_{H^+}$ of the
two--Higgs--doublet model (2HDM) of type II \cite{btau}. 
Here $\tan\beta$ is the ratio of the two Higgs vacuum expectation values and $M_{H^+}$
is the mass of the physical charged Higgs boson $H^+$ in the model.
Since the couplings of $H^+$ to $b$'s and $\tau$'s grow with
$\tan\beta$, $\BTNp$ probes large values of $\tan\beta$.
Earlier (but less powerful) constraints on the 2HDM were obtained by the OPAL
collaboration, which found $\tan\beta/M_{H^+}< 0.53\gev{}^{-1}$ from 
${\cal B}(\bar{B}\to X\tau\bar{\nu}_{\tau})$ \cite{opal1} 
and $\tan\beta/M_{H^+}< 0.78\gev{}^{-1}$ from       
${\cal B}(\tau \to \mu \bar{\nu}_{\mu} \nu_{\tau})$ \cite{opal2} at the 95\% CL. The direct search for 
a charged Higgs boson through $t \to b H^+$ at the Tevatron has yielded
slightly stronger bounds: 
$M_{H^+}> 125\gev$ for $\tan \beta=50$ and   
$M_{H^+}> 150\gev$ for $\tan \beta=70$ \cite{tev}.
In the low and intermediate $\tan\beta$ regions, the most constraining bound
currently comes from the FCNC--induced process
$b\to s\gamma$, which yields
$M_{H^+}> 295\gev$ independently of $\tan \beta$ \cite{bsg}.
At tree--level the Higgs sector of the Minimal Supersymmetric Standard Model (MSSM)
coincides with the type--II 2HDM.
The coupling of $H^+$ to fermions can be modified by a 
factor of order one due to $\tan\beta$--enhanced radiative 
corrections \cite{ar03,ltb},
yet this introduces only a few additional supersymmetric parameters
and the access to the Higgs sector in (semi-)leptonic $B$ decays is not obfuscated like in many other modes,
such as the loop--induced $b\to s\gamma$ decay.
This explains the great theoretical
interest in the experimental ranges for ${\cal B} (\BTNp)$~\cite{bpexp}.

The decay $\BDTN$ provides an alternative route to charged--Higgs
effects
%\cite{t,itoh,gh,ks,miki02,cg06,km08}
[9--15]. As we will show in the following,  
this mode is not only competitive with $B^+\to \tau^+\nu_\tau$, but also 
opens the door to a potential CP--violating phase in the Yukawa couplings
of the $H^+$ to $b$ and $\tau$. $\BDTN$ compares to $\BTN$ as follows:

\vspace{1mm}
i) ${\cal B} (\BDTN)$ exceeds ${\cal B} (\BTN)$
by roughly a factor of 50 in the Standard Model.

\vspace{1mm}
ii) $\BDTN$ involves the well--known element 
    $V_{cb}$ of the Cabibbo--Kobayashi--Maskawa (CKM) matrix. The 
    uncertainty on $|V_{ub}|$ entering $\BTN$ is much
    larger.

\vspace{1mm}
iii) ${\cal B} (\BTN)$ is proportional to two
    powers of the $B$ decay constant $f_B$, which must be obtained with
    non--perturbative methods. Current lattice gauge theory computations
    are struggling with chiral logarithms and $f_B^2$ can only be
    determined with an uncertainty of 30\% or more \cite{latt07}.
    $\BDTN$ involves two form factors, one of which can be
    measured in $\BDLN$ ($\ell=e,\mu$) decays \cite{b01,HFAG}. The
    other one is tightly constrained by Heavy Quark Effective
    Theory (HQET)
%\cite{HQET,n94,bgl,cn}
[19--22], so that hadronic uncertainties 
    can be reduced to well below 10\% once the measurement of $\BDLN$ is improved.

\vspace{1mm}
iv) Unlike $\BTN$ the three--body decay 
    $\BDTN$ permits the study of decay distributions 
    which discriminate between $W^+$ and $H^+$ exchange \cite{gh,ks,miki02}. 

\vspace{1mm}
v) The Standard Model (SM) contribution to $\BTN$
    is (mildly) helicity--suppressed, so that the sensitivity of ${\cal B}
    (\BTN)$ to $H^+$ is enhanced. For $\BDTN$ a similar effect only
    occurs near the kinematic endpoint, where the $D$ moves slowly
    in the $B$ rest frame \cite{gh}:
    While the transversely polarized $W^+$ contribution suffers
    from a P--wave suppression, the virtual $H^+$ recoils against the 
    $D$ meson in an unsuppressed S--wave.

\vspace{1mm}
Items iv) and v) strongly suggest to study differential decay
distributions in $\BDTN$. The $\tau$ in the final state
poses an experimental challenge, because it does not travel far enough
for a displaced vertex and its decay involves at least one more neutrino.
In particular, the $\tau$ polarization, known as a charged-Higgs
analyzer \cite{t}, is not directly accessible to experiment.
To our knowledge, the only theory papers which address the question of the missing information
on the $\tau$ momentum are \cite{gh,ks}, where a study of the $D$ meson energy spectrum is proposed.
Another straightforward way to deal with the missing information on the $\tau$ kinematics,
which in addition retains information on the $\tau$ polarization, is to consider the full
decay chain down to the detectable particles stemming from the $\tau$. 
We have studied the decays $\tau^- \to \pi^- \nu_\tau$, $\tau^- \to
\rho^- \nu_\tau$, and $\tau^- \to \ell^-\bar{\nu}_\ell \nu_\tau$ and
assessed the sensitivity of the decay distributions to $H^+$ effects.
We find that the decay chain $\bar{B}\to D \bar{\nu}_{\tau}\tau^-[\to \pi^-
\nu_{\tau}]$ discriminates between $W^+$ and $H^+$ exchange in an
excellent way. In this Letter we present the results for the
  differential decay rate as a function of the $D$ and $\pi^-$ energies
  and the angle between the $D$ and $\pi^-$ three--momenta for this decay
chain. Our result greatly facilitates the determination of
the effective coupling $g_S$ governing $H^+$ exchange,
including a potential complex phase, if e.g. a maximum likelihood fit
of the data to the theoretical decay distribution given below is employed.
A conventional analysis combining Monte Carlo simulations of $\bar{B}\to D
\bar{\nu}_{\tau} \tau^-$ and $\tau^- \to \pi^- \nu_\tau$ decays would be
very cumbersome, because the $\BDTN$ differential distributions strongly
depend on the a--priori unknown value~of~$g_S$.

\boldmath
\section{$B\rightarrow D$ Form Factors}
\unboldmath
The effective hamiltonian describing $B\rightarrow(D)\tau\nu_{\tau}$
transitions mediated by $W^+$ or $H^+$ reads (with $q=u\,(c)$) 
\begin{equation}
\label{eq:1}
\begin{split}
H_{\textrm{eff}} & =\frac{G_{F}}{\sqrt{2}} V_{qb}\text{\thinspace}
\big\{\left[  \overline{q}\gamma^{\mu}(1-\gamma_{5})b\right]  \text{\thinspace}
\left[  \overline{\tau}\gamma_{\mu}(1-\gamma_{5})\nu_{\tau}\right] \\
& -\frac{\ov m_{b}m_{\tau}}{m_{B}^{2}}\text{\thinspace}\text{\thinspace}
\overline{q}\left[ g_{S} + g_P \gamma_5\right] b  \text{\thinspace}
\left[\overline{\tau}(1-\gamma_{5})\nu_{\tau}\right]\big\}\,+\,\mbox{h.c.}
\end{split}
\end{equation}
The effective coupling constant $g_P$ only enters the $\BTN$ decay, while $\BDTN$ is only sensitive to $g_S\,$.
The $B^+$ meson mass $m_B$ is introduced in Eq.~(\ref{eq:1}), so that 
${\cal B} (\BTN)$ vanishes for $g_P=1$.
The above operators as well as $\ov m_b$ are defined in the $\ov{\rm MS}$ scheme.
In the MSSM, which is our main focus, one has $g_S=g_P$.

The analysis of $\BDTN$ requires the knowledge of the form factors $F_V$
and $F_S$ which parametrize the vector and scalar current matrix
elements:
\begin{align}% 
  \langle D(p_D)|\bar{c}\gamma^{\mu}b|\bar{B}(p_B)\rangle \! =\,&
  F_V(q^2)\! 
  \left[p_B^{\mu}+p_D^{\mu}-m_B^2\frac{1-r^2}{q^2}q^{\mu}\right]\nn
  +\,& F_S(q^2)\,m_B^2\frac{1-r^2}{q^2}q^{\mu}\,,\nn \langle
  D(p_D)|\bar{c}b|\bar{B}(p_B)\rangle=\,& \frac{m_B^2\,(1-r^2)}{\ov
    m_b-\ov m_c}F_S(q^2)\,,\label{eq:2}
\end{align}
where $p_B$ and $p_D$ denote the meson four--momenta, $q=p_B-p_D$, and $r=m_D/m_B\,$.
It is convenient to introduce the normalized form factors
$V_1\equiv F_V\,2\sqrt{r}/(1+r)$ and $S_1\equiv F_S(1+r)/(2\sqrt{r})$,
as well as the kinematic variable
\begin{equation}
\label{eq:3}
w\equiv(1+r^2-q^2/m_B^2)/2r.
\end{equation}
In the limit of infinitely heavy quark masses $m_Q=m_b,\,m_c$
(which are properly infrared--subtracted pole masses),
both $V_1(w)$ and $S_1(w)$ reduce to the universal
Isgur--Wise function $\xi(w)$, normalized to $\xi(1)=1$\,.
At the kinematic endpoint $w=1$, corrections to this limit read
\begin{equation}
\label{eq:4}
V_1(1) =\eta_v-\frac{1-r}{1+r}\big(\delta_{\textrm{rad}}+\delta_{1/m_Q}\big)\,,\quad
S_1(1) =\eta_v\,,
\end{equation}
up to $\mathcal{O}(\alpha_s^2,1/m_Q^2)$.
Here $\eta_v$ denotes radiative corrections in the limit of equal
heavy meson masses, and $\delta_{\textrm{rad}}\,(\delta_{1/m_Q})$ are the
first order radiative $(1/m_Q)$ corrections to the function $\xi_-$
defined in \cite{n94}.  The $\delta_{1/m_Q}$ term depends on the
  subleading function $\xi_3(w=1)=\bar{\Lambda}\eta(w=1)$ and on the HQET parameter $\bar{\Lambda}$. We take $\bar{\Lambda}=0.5\pm
0.1\,\textrm{GeV}$, $\eta(1)=0.6\pm 0.2$ \cite{lnn}, $\eta_v$ and
$\delta_{\textrm{rad}}$ to $\mathcal{O}(\alpha_s)$ from Ref.~\cite{n92},
and add a $5\%$ error to the form factors at $w=1$ to account for higher
order corrections. We obtain $V_1(1)=1.05\pm0.08$ and
$S_1(1)=1.02\pm0.05$.

\begin{figure}[t!]
{\psfrag{ix}{\raisebox{0.05cm}{\hspace{-0.3cm}\scalebox{1}{$w$}}}
\psfrag{iy}{\raisebox{0cm}{\hspace{-0.5cm}\scalebox{1}{$|V_{cb}|V_1(w)$}}}
\psfrag{1.0}{\hspace{-0.05cm}$\scalebox{0.8}{1.0}$}\psfrag{1.1}{\hspace{-0.05cm}$\scalebox{0.8}{1.1}$}\psfrag{1.2}{\hspace{-0.05cm}$\scalebox{0.8}{1.2}$}\psfrag{1.3}{\hspace{-0.05cm}$\scalebox{0.8}{1.3}$}\psfrag{1.4}{\hspace{-0.05cm}$\scalebox{0.8}{1.4}$}
\psfrag{1.5}{\hspace{-0.05cm}$\scalebox{0.8}{1.5}$}\psfrag{1.6}{\hspace{-0.05cm}$\scalebox{0.8}{1.6}$}
\psfrag{0.02}{\hspace{0.02cm}$\scalebox{0.8}{0.02}$}\psfrag{0.03}{\hspace{0.02cm}$\scalebox{0.8}{0.03}$}\psfrag{0.04}{\hspace{0.02cm}$\scalebox{0.8}{0.04}$}\psfrag{0.05}{\hspace{0.02cm}$\scalebox{0.8}{0.05}$}\psfrag{0.06}{\hspace{0.02cm}$\scalebox{0.8}{0.06}$}\psfrag{0.07}{\hspace{0.02cm}$\scalebox{0.8}{0.07}$}
\begin{center}
\hspace{-0.2cm}
\scalebox{0.8}{\epsfig{file=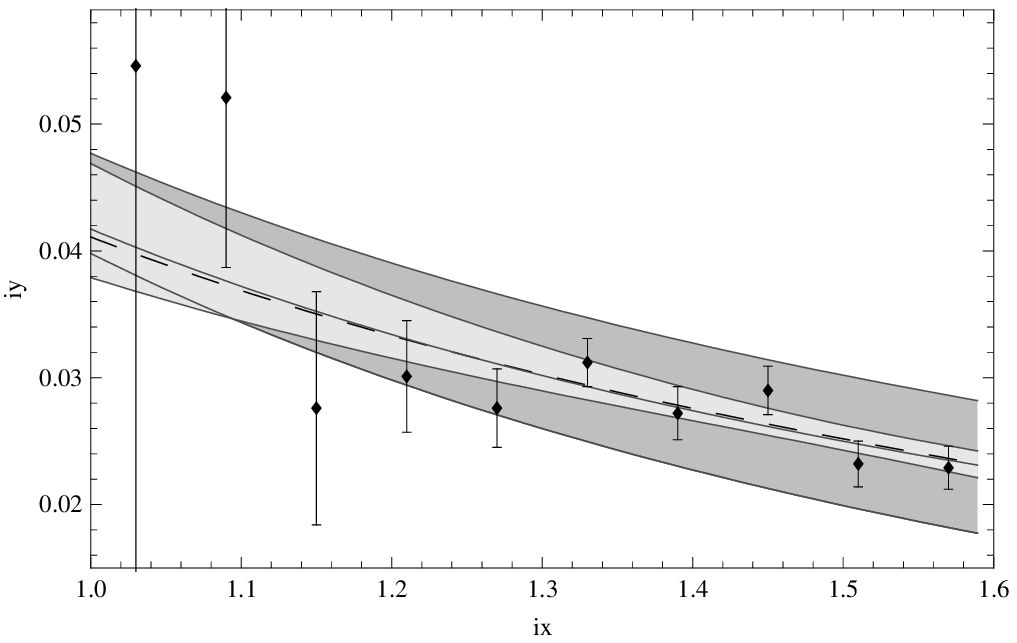}}
\end{center}
}
\vspace{-0.8cm}
\caption{Vector form factor $V_1(w)$. Dots: exp. data \cite{b01} with
  stat. errors only. Dashed: fit to parametrization in \cite{cn}. Plain:
  fit to linear parametrization in \cite{rh}. Dark gray band: form factor with
  HQET constraint at $w=1$; systematic errors dominate at large
  recoil. Light gray band: form factor from HFAG \cite{HFAG}.}
\label{ff}
\end{figure}

The semileptonic decay into light leptons $\BDLN$ depends solely on the
vector form factor $V_1(w)$. The measured quantity $|V_{cb}|V_1(w)$ was
fitted by the BELLE collaboration \cite{b01} to a two--parameter ansatz
$V_1(w,V_1(1),\rho_1^2)$ \cite{cn} derived from dispersion relations and
heavy quark spin symmetry \cite{bgl}.  The fitted curve, however,
suffers from large statistical and systematic uncertainties:
$|V_{cb}|V_1(1)=(4.11\pm0.44\pm0.52)\%$, $\rho_1^2=1.12\pm0.22\pm0.14$
\cite{b01}.  We thus take $V_1(1)$ from HQET instead, use
$|V_{cb}|=(4.17\pm0.07)\%$ from inclusive semileptonic $B$ decays
\cite{pdg}, and only fix the form factor at large recoil $w=1.45$ from
the data, including the dominant systematic errors in a conservative
way: $|V_{cb}|V_1(1.45)=(2.63\pm0.51)\%$.  The form factor over the
whole kinematic range is then obtained using a two--parameter
description $F_{V}(w,a_0^{V},a_1^{V})$, which uses a conformal mapping
$w\to z(w)$ resulting in an essentially linear dependence of $F_V$ on
$z$ \cite{rh}. This linearity in $z(w)$ is confirmed by the fact that
fitting the $\BDLN$ data with both $F_V$ parametrizations without
further theoretical constraints essentially gives the same result (see
Fig.~\ref{ff}).  The sets of parameters corresponding to the minimal and
maximal form factors satisfying the HQET constraint at $w=1$ are
displayed in Tab.~\ref{par} for both parametrizations
$V_1(w,V_1(1),\rho_1^2)$ and $F_V(w,a_0^V,a_1^V)$.  They delimit the
dark gray area in Fig.~\ref{ff}.  We stress that the large error band in
Fig.~\ref{ff} at large $w$ is not due to theory uncertainties but rather
to the large systematic error on $|V_{cb}|V_1(1.45)$ from \cite{b01}.

We choose to use only the most recent set of experimental data for our numerical analysis.
The HFAG \cite{HFAG} treats systematic errors in a different way and, including the older CLEO and ALEPH data,
finds smaller uncertainties at large recoil (see light gray band in Fig.~\ref{ff}.
The corresponding minimal and maximal curves are given in good approximation
by the parameters in the first two lines of Tab.~\ref{tab:HFAG} for $w$ inside the $\BDTN$ phase space).
The vector form factor has also been studied on the lattice.
Computations with quenched Wilson \cite{FvQuenched} and dynamical staggered \cite{FvStaggered} fermions,
however, both suffer from potentially large systematic errors, which are not fully controlled.
In the end, the improvements in the measurements of the $\BDLN$ and $\BDTN$ modes will go together,
and $|V_{cb}|F_V$ will most likely be best determined from experimental data alone.
For the time being, we will proceed with the conservative estimation of Tab.~\ref{par}.

\begin{table}[t!]
\begin{center}
\begin{tabular}{|l|c|c|c|}
\hline
Parameters & min. $|V_{cb}|F$ & max. $|V_{cb}|F$ & centr. $|V_{cb}|F$\\
\hline
%$\{V_1(1),\rho_1^2\}$ & $\{0.97,1.47\}$ & $\{1.13,1.06\}$ & $\{1.05,1.24\}$\\
%$\{a_0^V,a_1^V\}[10^{-4}]$ & $\{2.30,-14\}$ & $\{3.01,-5\}$ & $\{2.66,-9\}$\\
%$\{a_0^S,a_1^S\}[10^{-3}]$ & $\{3.96,-26\}$ & $\{5.05,-7\}$ & $\{4.52,-16\}$\\
$\{|V_{cb}|V_1(1),\rho_1^2\}$ & $\{0.040,1.47\}$ & $\{0.048,1.06\}$ & $\{0.044,1.24\}$\\
$|V_{cb}|\{a_0^V,a_1^V\}[10^{-5}]$ & $\{0.94,-5.7\}$ & $\{1.28,-2.2\}$ & $\{1.11,-3.9\}$\\
$|V_{cb}|\{a_0^S,a_1^S\}[10^{-4}]$ & $\{1.62,-1.1\}$ & $\{2.14,-3.2\}$ & $\{1.88,-6.8\}$\\
\hline
\end{tabular}
\end{center}
\vspace{-0.5cm}
\caption{Parameters $\{|V_{cb}|V_1(1),\rho_1^2\}$ for $|V_{cb}|F_V$ \cite{cn} and $\{|V_{cb}|a_0^{V,S},|V_{cb}|a_1^{V,S}\}$ for $|V_{cb}|F_{V,S}$ (see
  \cite{rh}, $Q^2=0$, $\eta=2$, subthreshold poles:
  $m(1^-)=6.337,\,6.899,\,7.012\,\textrm{GeV}$ and
  $m(0^+)=6.700,\,7.108\,\textrm{GeV}$ \cite{poles}).
$F_V$ is displayed in dark gray in Fig.~\ref{ff}.}
\label{par}
\end{table}

\begin{table}[t!]
\begin{center}
\begin{tabular}{|l|c|c|c|}
\hline
Parameters & min. $|V_{cb}|F$ & max. $|V_{cb}|F$ & centr. $|V_{cb}|F$\\
\hline
$\{|V_{cb}|V_1(1),\rho_1^2\}$ & $\{0.038,1.01\}$ & $\{0.047,1.30\}$ & $\{0.042,1.17\}$\\
$|V_{cb}|\{a_0^V,a_1^V\}[10^{-5}]$ & $\{1.03,-1.3\}$ & $\{1.17,-4.8\}$ & $\{1.10,-3.0\}$\\
$|V_{cb}|\{a_0^S,a_1^S\}[10^{-4}]$ & $\{1.78,-5.7\}$ & $\{2.00,-7.6\}$ & $\{1.89,-6.6\}$\\
\hline
\end{tabular}
\end{center}
\vspace{-0.5cm}
\caption{Parameters $\{|V_{cb}|V_1(1),\rho_1^2\}$ for $|V_{cb}|F_V$ from HFAG \cite{HFAG},
and $\{|V_{cb}|a_0^{V,S},|V_{cb}|a_1^{V,S}\}$ for $|V_{cb}|F_{V,S}$.
$F_V$ is displayed in light gray in Fig.~\ref{ff}.}
\label{tab:HFAG}
\end{table}

In a similar way, the scalar form factor $F_{S}(w,a_0^{S},a_1^{S})$ is
constrained by HQET at $w=1$, while its value at large recoil is fixed
from the relation $F_S(q^2=0)=F_V(q^2=0)$.  The resulting parameters are
displayed in the third line of Tab.~\ref{par} (or Tab.~\ref{tab:HFAG} if
$F_V$ is taken from \cite{HFAG}).  As expected from the heavy-quark
limit, the normalized form factor $S_1$ is quite close to $V_1$ on the
whole $w$ range, with slightly smaller errors.

\section{Charged--Higgs Effects}
The MSSM is a well--motivated new--physics scenario in which
charged scalar current interactions occur at tree--level.
Resumming the dominant $\tan\beta$-enhanced loop corrections to all
orders, the couplings $g_{S,P}$ in Eq.~(\ref{eq:1}) specify to \cite{yuk,itoh}
\begin{equation}
\label{eq:5}
g_{S}=g_{P}=\frac{m_{B}^{2}}{M_{H^+}^{2}}\, 
\frac{\tan^{2}\beta}{ (1+\widetilde \epsilon_0\,\tan\beta)
                      (1+\epsilon_{\tau}\,\tan\beta)}\,.
\end{equation}
This particular form holds in MSSM scenarios with Minimal Flavor 
Violation (MFV).
The loop factor $\widetilde \epsilon_0$ arises from the quark Yukawa sector
and depends on ratios of superparticle masses, resulting in a
  sizable non--decoupling effect $\widetilde
\epsilon_0\tan\beta={\cal O}(1)$ for $\tan\beta={\cal O}(50)$.  $\epsilon_{\tau}$
comprises the corresponding effect for the $\tau$ lepton. $\widetilde
\epsilon_0$ and $\epsilon_\tau$ can receive sizable complex phases from
the Higgsino mass parameter $\mu$, if first--generation sfermions are
sufficiently heavy to soften the impact of the bounds on electric dipole moments on
$\arg \mu$. Beyond MFV also phases from squark mass matrices will easily
render $g_S$ complex.
It is therefore mandatory to constrain -- and eventually measure -- both
magnitude and phase of $g_S$. The type--II 2HDM is recovered
by setting $\widetilde \epsilon_0=\epsilon_{\tau}=0$.

\begin{figure*}[t!]
\begin{minipage}{0.5\linewidth}
{\psfrag{ix}{\raisebox{0.1cm}{\hspace{-0.2cm}$g_P$}}
\psfrag{iy}{\raisebox{0cm}{\hspace{-1.2cm}$\mathcal{B}(B\to\tau\nu)\,[10^{-3}]$}}
\psfrag{X1}{$1\sigma$}
\psfrag{X2}{$3\sigma$}
\begin{center}
\hspace{-1.5cm}
\scalebox{0.9}{\epsfig{file=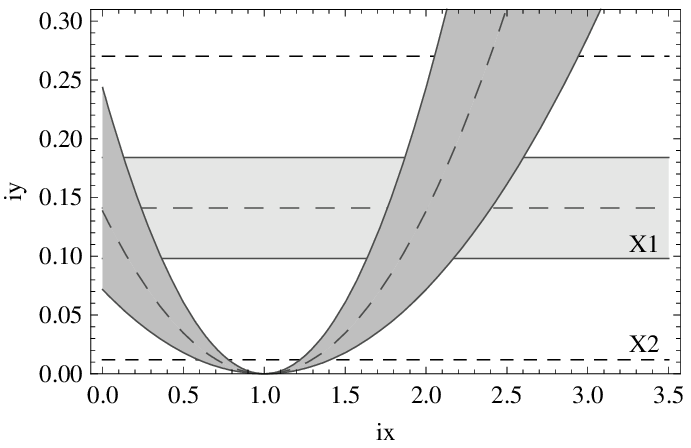}}
\end{center}
}
\end{minipage}
\begin{minipage}{0.45\linewidth}
{\psfrag{ix}{\raisebox{0.1cm}{$g_S$}}
\psfrag{iy}{\raisebox{-0.2cm}{\hspace{-0.3cm}$R$}}
\psfrag{X1}{$1\sigma$}
\psfrag{0.15}{}\psfrag{0.25}{}\psfrag{0.35}{}\psfrag{0.45}{}
\psfrag{0.0}{\hspace{0.06cm}$\scalebox{0.8}{0}$}\psfrag{0.5}{\hspace{-0.04cm}$\scalebox{0.8}{0.5}$}\psfrag{1.0}{\hspace{0.06cm}$\scalebox{0.8}{1}$}\psfrag{1.5}{\hspace{-0.04cm}$\scalebox{0.8}{1.5}$}\psfrag{2.0}{\hspace{0.06cm}$\scalebox{0.8}{2}$}\psfrag{2.5}{\hspace{-0.04cm}$\scalebox{0.8}{2.5}$}\psfrag{3.0}{\hspace{0.06cm}$\scalebox{0.8}{3}$}\psfrag{3.5}{\hspace{-0.04cm}$\scalebox{0.8}{3.5}$}
\psfrag{0.10}{\hspace{0.12cm}$\scalebox{0.8}{0.1}$}\psfrag{0.20}{\hspace{0.12cm}$\scalebox{0.8}{0.2}$}\psfrag{0.30}{\hspace{0.12cm}$\scalebox{0.8}{0.3}$}\psfrag{0.40}{\hspace{0.12cm}$\scalebox{0.8}{0.4}$}
\begin{center}
\hspace{-0.2cm}
\scalebox{0.8}{\epsfig{file=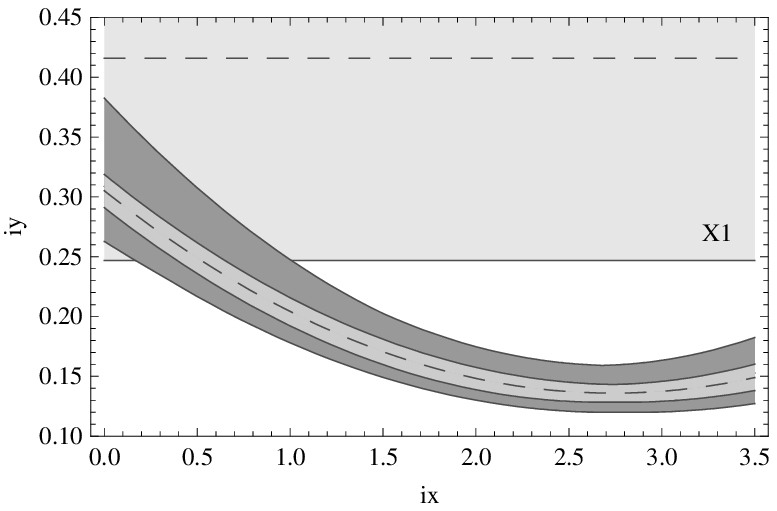}}
\end{center}
}
\end{minipage}
\vspace{-0.4cm}
\caption{Left: $\mathcal{B}(B\to\tau\nu)$ as a function of $g_P$. Light
  gray band: $\mathcal{B}^{\textrm{exp}}=(1.41\pm0.43)\times10^{-4}$ \cite{bpexp}. Gray band: $\mathcal{B}^{\textrm{th}}$,
  $|V_{ub}|=(3.86\pm0.09\pm0.47)\times10^{-3}$ \cite{vub}, main error from
  $B$ decay
  constant $f_B=(216\pm38)\textrm{MeV}$ \cite{latt07}. Right:
  $R\equiv\mathcal{B}(\BDTN)/\mathcal{B}(\BDLN)$ as a function of $g_S$.
Light gray band: $R^{\textrm{exp}}$, see
  (\ref{eq:6}) \cite{babarbdt}. Dark gray band: $R^{\textrm{th}}$, see (\ref{eq:7}).
Gray band: $R^{\textrm{th}}$ with the HFAG vector form factor, see Tab.~\ref{tab:HFAG}.
SM:
$R^{\textrm{th}}(g_S=0)=0.31\,{}^{+0.07}_{-0.05}\,(\textrm{dark gray})\,[\,\pm 0.02\,(\textrm{gray})]$.}
\label{bdt}
\end{figure*}

The $\BDTN$ branching ratio has recently been measured by the BABAR collaboration \cite{babarbdt}:
\begin{equation}
\label{eq:6}
R^{\textrm{exp}}\equiv\frac{\mathcal{B}(\BDTN)}{\mathcal{B}(\BDLN)}=(41.6\pm 11.7\pm 5.2)\%\,.
\end{equation}
The normalization to $\mathcal{B}(\BDLN)$ reduces the dependence
on the vector form factor $F_V$ and thus tames the main theoretical uncertainties.
In the presence of charged--Higgs contributions, the theoretical ratio is approximated to 1\% by
\begin{equation*}
R^{\textrm{th}}=\frac{ 1.126 + 0.037\, r_V + r_0^2\, ( 1.544 + 0.082\, r_S + N_{H^+} ) }{ 10 - 0.95\, r_V }\ ,
\end{equation*}
\vspace{-5mm}
\begin{equation}
\label{eq:7}
\begin{split}
N_{H^+}=&- r_{cb}\,Re[g_S]\,( 1.038 + 0.076\, r_S )\\ 
        &+ r_{cb}^2\,|g_S|^2\,( 0.186 + 0.017\, r_S ),
\end{split}
\end{equation}
with $r_V=(a_1^V/a_0^V)/(-3.4)$, $r_S=(a_1^S/a_0^S)/(-3.5)$,
$r_0=(a_0^S/a_0^V)/17$, and $r_{cb}=0.8/(1-\ov m_c/\ov m_b)$.  The
dependence on the slope parameters $a_1^{V,S}$ appears to be quite mild.
In Fig.~\ref{bdt} we compare $R^{\textrm{th}}$ (right--hand side) as
well as $\mathcal{B}(B\to\tau\nu)$ (left--hand side) to their one--sigma
measurements for positive $g_S$ and $g_P$.  For $R^{\textrm{th}}$, we
also display the less conservative theoretical prediction obtained from
the HFAG vector form factor in Tab.~\ref{tab:HFAG} (light gray band). In
particular, we obtain the SM estimates 
\begin{displaymath}
{\cal B} (B^- \to D^0 \tau^- \bar{\nu}_{\tau})^{\textrm{SM}} 
   = (0.71\pm0.09)\% 
\end{displaymath}
and 
\begin{displaymath}
{\cal B} (\bar{B}^0 \to D^+ \tau^- \bar{\nu}_{\tau})^{\textrm{SM}} 
   = (0.66\pm 0.08)\%. 
\end{displaymath}
(Error sources: $|V_{cb}|F_V(w),\ S_1(1),\ |V_{cb}|$). We cannot
reproduce the small errors of Ref.~\cite{cg06}.

The $\BDTN$ branching fraction is promising to discover -- or constrain --
charged--Higgs effects, but not to measure $g_S$ with good precision, as the dependence in Fig.~\ref{bdt} is too flat.
The differential distribution in the decay chain
$\bar{B}\to D \bar{\nu}_{\tau}\tau^-[\to\pi^-\nu_{\tau}]$ is better
suited for that purpose.
The experimentally accessible quantities are the energies $E_D$ and
$E_\pi$ of the $D$ and $\pi^-$ mesons, respectively, and the angle
$\theta$ between the three--momenta $\vec p_D$ and $\vec p_\pi$. We
define these quantities in the $B$ rest frame,
which can be accessed from the $\Upsilon(4S)$ rest frame thanks to full $B$ reconstruction \cite{babarbdt}.
We integrate over the phase space of the two unobserved
neutrinos in the final state. Our formulae contain the
full spin correlation between the production and decay of the $\tau$,
which is important to discriminate between SM and charged--Higgs contributions.
This approach further facilitates the rejection of
backgrounds from neutral particles escaping detection, as in $\bar{B}\to D
D^-[\to \pi^- \pi^0]$ with an undetected $\pi^0$: If the mass of the
undetected particle is $m$, this background can be suppressed by cuts
excluding the region around
\begin{align}
\cos\theta = \frac{(m_B-E_D-E_{\pi})^2-2(E_D^2-m_D^2)-m^2}{2(E_D^2-m_D^2)}\,. \label{eq:8}
\end{align}

We obtain the differential distribution
\begin{align}
  & \frac{d\Gamma(\bar{B}\to D\bar{\nu}_{\tau}\tau^-
    [\to\pi^-\nu_{\tau}])}{dE_D\,dE_{\pi}\,d\cos\theta}=G_F^4f_{\pi}^2|V_{ud}|^2|V_{cb}|^2\tau_{\tau}\label{eq:9}\\
  &
  \times\left[C_W(F_V,F_S)-C_{WH}(F_V,F_S)\,Re[g_S]+C_H(F_S)|g_S|^2\right]\nonumber
\end{align}
with form--factor--dependent functions of $E_D$, $E_{\pi}$, and
$\cos\theta$ for the SM ($C_W$), interference
($C_{WH}$), and Higgs ($C_H$) contributions, given 
as follows for vanishing $m_\pi$ (this approximation, 
which is good to 1\%, is not used in our numerical analysis),
\begin{equation}
\begin{split}
C_W& =\kappa \frac{m_{\tau}^4}{2}\frac{l^2}{p_{\pi}\cdot
    l} \, 
  \bigg\{ P^2 (b-1) + (P \cdot l)^2 \frac{2b}{l^2} + \\
  & \Bigg[
   \frac{l^2 (P\cdot p_\pi)^2 }{ (p_{\pi}\cdot l)^2} 
   - \frac{2 (P\cdot l)(P\cdot p_{\pi}) 
         }{p_{\pi}\cdot l} \Bigg] (3 b - 1)
  \bigg\}\, , \\
C_{WH} & = 2\kappa \, m_{\tau}^4  \, 
\frac{(1-r^2)F_S}{1-\ov m_c/\ov m_b} \, b \, 
  \bigg[ P \cdot l -\frac{l^2 P\cdot p_\pi }{p_{\pi}\cdot l} \bigg]\, , \\
C_H &=\kappa\,m_{\tau}^6 \frac{(1-r^2)^2
    F_S^2}{(1-\ov m_c/\ov m_b)^2}
  \Big(1-\frac{m_{\tau}^2}{2\,p_{\pi}\cdot l} \Big)\, , \\
\end{split} \label{defc}
\end{equation}
where $\ov m_c$ and $\ov m_b$ must be evaluated at the same scale so that 
$\ov m_c/\ov m_b=0.20\pm 0.02$ \cite{run}, and   
\begin{eqnarray}
P \!\!& =&\! F_V (p_B+p_D) - 
    (F_V-F_S)\frac{m_B^2(1-r^2)}{q^2} (p_B-p_D) \,, \nn
\kappa \!\!& =&\!
\frac{E_{\pi}\sqrt{E_D^2-m_D^2}}{
             128\,\pi^4\,m_B m_{\tau}}\,,
\quad b = \frac{m_\tau^2}{p_\pi \cdot l} 
      \Big( 1 - \frac{m_{\tau}^2}{2\,p_{\pi}\cdot l} \Big)\,, \nn
l \!\!& =&\! p_B-p_D-p_\pi \,,\quad 
   q^2 = (p_B-p_D)^2.\label{defpkb}
\end{eqnarray}
The dot products appearing in \eqsand{defc}{defpkb} are related to the energies,
momenta, and the angle $\theta$ measured in the $B$ rest frame as
$p_B\!\cdot\! l=m_B (m_B-E_D-E_\pi)$,
$p_D\!\cdot\! l = E_D (m_B -E_D-E_\pi) +
|\vec p_D|^2+|\vec p_D| E_\pi \cos\theta$, 
 $p_\pi\!\cdot\! l = E_\pi (m_B-E_D) 
+ |\vec p_D| E_\pi \cos\theta$, and $p_B\!\cdot\! p_D = m_B E_D$.
 Further $\tau_{\tau}=(290.6\pm 1.0)\times 10^{-15}s$ is the $\tau$
lepton lifetime, $f_{\pi}=(130.7\pm 0.1\pm 0.36)\,\textrm{MeV}$ the
pion decay constant, and the CKM matrix elements are $|V_{ud}|=0.97377\pm
0.00027$ and $|V_{cb}|=(41.7\pm 0.7)\times 10^{-3}$, the latter being
well determined from inclusive semileptonic $B$ decays \cite{pdg}.
Remarkably, one can probe a CP--violating phase of $g_S$ by exploiting the
shape of the distribution in Eq.~(\ref{eq:9}), which is not possible from the branching
fraction of either $\BDTN$ or $\BTN$.

\begin{figure*}[t!]
\begin{minipage}{0.5\linewidth}
{\psfrag{ix}{$\cos\theta$}
\psfrag{iy}{\hspace{-2.2cm}$d\Gamma(B^0\to D^+\pi^-\bar{\nu}\nu)\
  [10^{-16}\textrm{GeV}]$}
\psfrag{X1}{$E_D=2\,\textrm{GeV}$, $E_{\pi}=1\,\textrm{GeV}$}
\psfrag{X2}{$g_S=0$ (SM)}
\psfrag{X3}{$g_S=1+i$}
\psfrag{X4}{$g_S=2$}
\begin{center}
\hspace{-1.5cm}
\scalebox{0.8}{\epsfig{file=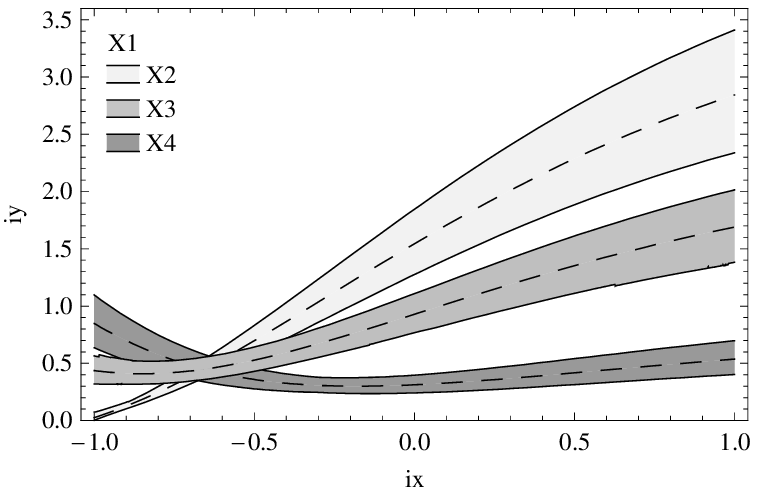}}
\end{center}
}
\end{minipage}
\begin{minipage}{0.45\linewidth}
{\psfrag{ix}{$\cos\theta$}
\psfrag{iy}{\hspace{-2.2cm}$d\Gamma(B^0\to D^+\pi^-\bar{\nu}\nu)\
  [10^{-16}\textrm{GeV}]$}
\psfrag{X1}{$E_D=2\,\textrm{GeV}$, $E_{\pi}=1\,\textrm{GeV}$}
\psfrag{X2}{$g_S=0$ (SM)}
\psfrag{X3}{$g_S=0.5$}
\begin{center}
\hspace{0cm}
\scalebox{0.8}{\epsfig{file=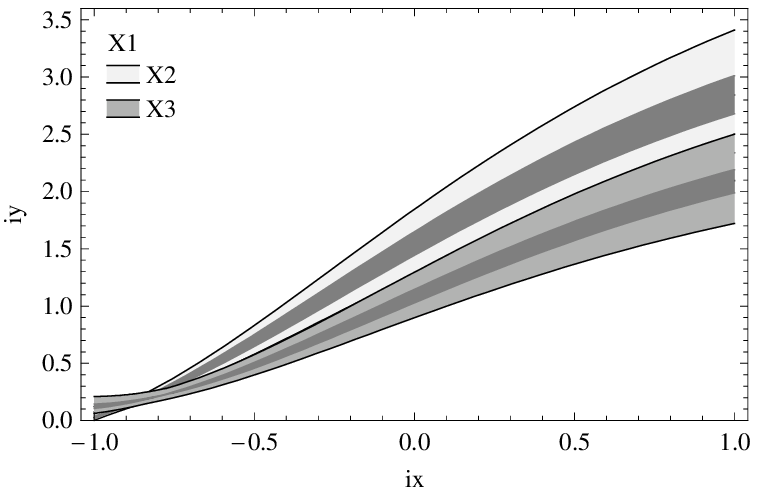}}
\end{center}
}
\end{minipage}
\vspace{-0.4cm}
\caption{
$\bar{B}^0\to D^+\bar{\nu}_{\tau}\tau^-[\to\pi^-\nu_{\tau}]$ angular distribution
for $E_D=2\gev$ and $E_{\pi}=1\gev$.
Left: $g_S=0,1+i,2$. Right: $g_S=0,0.5$ (dark gray: without uncertainties in $F_V(w)$ and $V_{cb}$,
errors from $S_1(1)$ and $\ov m_c/\ov m_b$). The conservative form factor estimates of Tab.~\ref{par} were considered.
\label{bdp}
}
\end{figure*}

For illustration, we show the differential decay distribution including
charged--Higgs effects in comparison with the SM for the meson energies
$E_D=2\,\textrm{GeV}$ and $E_{\pi}=1\,\textrm{GeV}$, so that the
whole range of $\cos\theta$ is kinematically accessible. In this
particular region of phase space the SM rate is
strongly suppressed for $\cos\theta=-1$. For a large scalar coupling
$g_S=2$ (Fig.~\ref{bdp}, left), the Higgs contribution dominates the rate at this
point (dark gray band), so that we can clearly distinguish it from
the SM (light gray band).
The experimental information from $\mathcal{B}(\BTN)$ constrains $|1-g_P|$. For real
$g_P$ this permits a range near $g_P=0$ and another range around
$g_P=2$. In the MSSM situation with $g_P=g_S$, the case $g_S=2$
therefore is in agreement with $\BTN$, but can be
confirmed or ruled out by measuring our distribution. The discrimination potential for the phase
of $g_S$ shows up in the light gray band: It corresponds to a complex
$g_S=1+i$, which yields the same
$\mathcal{B}(\BTN)$ as $g_S=0,2$.
The $\BDTN$ branching ratio alone may also help to distinguish between these solutions,
depending on the future experimental value of $\mathcal{B}(\BDTN)$, see Fig.~\ref{bdt}.
For general $g_S$ values, a fit to the triple differential distribution in
Eq.~(\ref{eq:9}) would excellently
quantify charged--Higgs effects, especially once better experimental information on the form factors is available,
as we illustrate with fixed $D$ and $\pi^-$ energies in Fig.~\ref{bdp} (right--hand side) for $g_S=0.5$.
Such a fit would combine information from different parts of the phase space,
and thus resolve much smaller $g_S$ values.
A more precise quantitative analysis would require the fit to actual data, and thus goes beyond the scope of this paper.
Still, keep in mind that even with more precise $\BTN$ experimental data and improved estimates
of $f_B$ and $|V_{ub}|$, a value of $g_P\simeq0.2-0.3$ will be very difficult to exclude with $\mathcal{B}(\BTN)$.
$\BDTN$ is thus definitely competitive.

As mentioned in the Introduction, a similar analysis was performed for
the other $\tau$ decay channels $\tau^-\to\rho^-\nu_\tau$ and
$\tau^-\to\ell^-\bar{\nu}_\ell \nu_\tau$, which together with
$\tau^-\to\pi^-\nu_\tau$ constitute more than 70\% of the $\tau$
branching fraction.  Ultimately, a combined analysis of all these modes
is desirable in order to exploit the available and forthcoming
experimental data in an optimal way.

\section{Conclusions}
We have studied charged--Higgs effects in a differential distribution of the
decay chain $\bar{B}\to D\bar{\nu}_{\tau}\tau^-[\to\pi^-\nu_{\tau}]$, 
which has the following advantages over the branching fractions
$\mathcal{B}(\BTN)$ and $\mathcal{B}(\BDTN)$:

\vspace{1mm} 
i) The Higgs coupling constant $g_S$ can be determined from 
  the \emph{shape} of the distribution in sensitive phase space regions.
  This analysis should be possible with current $B$ factory data.

\vspace{1mm}
ii) The dependence on both $|g_S|$ and $\real [g_S]$ allows to
  quantify a possible CP--violating phase. 
  Since our decay distribution is a 
  CP--conserving quantity, the phase of $g_S$ is determined with a  
  two--fold ambiguity. In the MSSM such a phase stems from 
  the $\mu$ parameter or the soft breaking terms and enters through 
  $\tan\beta$--enhanced loop factors. 
  $B\to D\tau \nu$ complements 
  collider studies of these phases \cite{ccp}.

\vspace{0.8mm} 
The main uncertainties stem from the form factors. One can gain a
much better accuracy with better data on the vector form factor
$F_V$. The recent $\BDLN$ measurement by BABAR
\cite{babarbdl} furnishes promising data for a new fit.

Within the MSSM, one will be able to place new constraints 
on the $\tan\beta-M_{H^+}$ plane, once our results are confronted with 
actual data from the $B$ factories. If $\tan\beta/M_{H^+}$ is indeed
large, there is a fair chance to reveal charged--Higgs effects ahead of
the LHC.
\vspace{3mm}\\
\emph{Acknowledgments}: The authors acknowledge helpful discussions
  with Richard Hill and Michael Feindt.  This work is supported in part
  by the DFG grant No.~NI 1105/1--1, by the DFG--SFB/TR9,
and by the EU Contract No.~MRTN-CT-2006-035482, \lq\lq FLAVIAnet''.


\vspace{-1mm}

\begin{thebibliography}{00}

% #1 
\bibitem{btau}
% first H+ in B -> tau nu
  W.~Hou,
  %``Enhanced charged Higgs boson effects in B- $\to$ tau anti-neutrino, mu
  %anti-neutrino and b $\to$ tau anti-neutrino + X,''
  Phys.\ Rev.\  D {\bf 48}, 2342 (1993).
  %%CITATION = PHRVA,D48,2342;%%
% recent B -> tau nu, refer to measurement
%  A.~Akeroyd and C.~Chen,
  %``Effect of H^\pm on B^\pm\to \tau^\pm\nu_\tau and D^\pm_s\to
  %                \mu^\pm\nu_\mu,\tau^\pm\nu_\tau''
%  Phys.\ Rev.\ D {\bf 75}, 075004 (2007)
%  [hep-ph/0701078].
  %%CITATION = HEP-PH/0701078;%%"

% #2 
\bibitem{opal1}
  OPAL Collaboration,
  %``Measurement of the branching ratio for the 
  %process b --> tau-  anti-nu/tau X,''
  Phys.\ Lett.\  B {\bf 520}, 1 (2001).
 % [arXiv:hep-ex/0108031]
  %%CITATION = PHLTA,B520,1;%%

% #3 
\bibitem{opal2}
 OPAL Collaboration,
  %``A measurement of the tau- --> mu- anti-nu/mu nu/tau branching ratio,''
  Phys.\ Lett.\  B {\bf 551}, 35 (2003).
 % [arXiv:hep-ex/0211066].
  %%CITATION = PHLTA,B551,35;%%

% #4 
\bibitem{tev}
 J.~Nielsen [CDF and D0 Collaborations],
  %``Tevatron searches for Higgs bosons beyond the standard model,''
  Nucl.\ Phys.\ Proc.\ Suppl.\  {\bf 177}, 224 (2008).
  %%CITATION = NUPHZ,177-178,224;%%

% #5 
\bibitem{bsg}
M.~Misiak \emph{et al.}, Phys. Rev. Lett. {\bf 98}, 022002 (2007).

% #6
% historical: susy-corr. to yukawa couplings
\bibitem{ltb}
 L.~Hall, R.~Rattazzi and U.~Sarid,
 %``The Top quark mass in supersymmetric SO(10) unification''
 Phys.\ Rev.\ D {\bf 50}, 7048 (1994);
 %%CITATION = HEP-PH/9306309;%%"
T.~Blazek, S.~Raby and S.~Pokorski,
 %``Finite supersymmetric threshold corrections to CKM matrix elements in the large tan Beta regime''
 Phys.\ Rev.\ D {\bf 52}, 4151 (1995).
% [hep-ph/9504364];
 %%CITATION = HEP-PH/9504364;%%"
M.~Carena \emph{et al.},
  %``Effective Lagrangian for the anti-t b H+ interaction in the MSSM and
  %charged Higgs phenomenology,''
  Nucl.\ Phys.\  B {\bf 577}, 88 (2000).
%  [hep-ph/9912516].
  %%CITATION = NUPHA,B577,88;%%
For earlier work on $t\to b H^+$ without resummation of $\tan\beta$--enhanced radiative corrections,
see R.A.~Jimenez and J.~Sola, Phys. Lett. B {\bf 389}, 53 (1996).

% #7
\bibitem{ar03}
 A.~Akeroyd and S.~Recksiegel, J. Phys. G {\bf 29}, 2311 (2003). 
% [hep-ph/0306037].

% #8 
\bibitem{bpexp}
  BELLE Collaboration,
  %``Evidence of the purely leptonic decay B- --> tau- anti-nu/tau''
  Phys.\ Rev.\ Lett.\ {\bf 97}, 251802 (2006);
  %%CITATION = HEP-EX/0604018;%%
  BABAR Collaboration,
  %``A search for B+ --> tau+ nu with Hadronic B tags''
  Phys.\ Rev.\ D {\bf 76}, 052002 (2007);
%  [0708.2260][hep-ex].
  %%CITATION = ARXIV:0708.2260;%%"
average: D.~Monorchio, talk presented at HEP2007,
http://www.\linebreak[2] hep.man.ac.uk/HEP2007/.

% #9
\bibitem{gh}
% first consideration of B -> D tau nu; diff. rate
  B.~Grzadkowski and W.~Hou,
  %``Searching for B $\to$ D tau anti-tau-neutrino at the 10-percent level,''
  Phys.\ Lett.\  B {\bf 283}, 427 (1992).
  %%CITATION = PHLTA,B283,427;%%

% #10
\bibitem{t}
% B->D tau nu total rate (HQET)
  M.~Tanaka,
  %``Charged Higgs effects on exclusive semitauonic B decays,''
  Z.\ Phys.\  C {\bf 67}, 321 (1995).
%  [hep-ph/9411405].
  %%CITATION = ZEPYA,C67,321;%%

% #11
\bibitem{ks}
% suggest calculation of decay chain B -> D tau -> final particles; diff. rate
  K.~Kiers and A.~Soni,
  %``Improving constraints on tan(beta/m(H)) using B --> D tau anti-nu,''
  Phys.\ Rev.\  D {\bf 56}, 5786 (1997).
%  [hep-ph/9706337].
  %%CITATION = PHRVA,D56,5786;%%

% #12
% include yukawa corrections in the MSSM; diff. rate
\bibitem{miki02}
 T.~Miki, T.~Miura and M.~Tanaka,
 %``Effects of charged Higgs boson and QCD corrections in anti-
 %                 B --> D tau anti-nu/tau''
 %%CITATION = HEP-PH/0210051,%%
 [hep-ph/0210051].

% #13
\bibitem{itoh}
%extensive study of yukawa-corrections in B -> D tau nu; total rate only
  H.~Itoh, S.~Komine and Y.~Okada,
  %``Tauonic B decays in the minimal supersymmetric standard model''
  Prog.\ Theor.\ Phys.\  {\bf 114}, 179 (2005).
%  [hep-ph/0409228].
  %%CITATION = PTPKA,114,179;%%

% #14
\bibitem{cg06}
C.-H.~Chen and C.-Q.~Geng, JHEP {\bf 0610}, 053 (2006).

% #15
\bibitem{km08}
J.F.~Kamenik and F.~Mescia, [0802.3790][hep-ph].

% #16
\bibitem{latt07} 
% lattice result for fB
 M.~Della Morte,
 %``Standard Model parameters and heavy quarks on the lattice''
 PoS\ LAT2007, 8.
% [0711.3160][hep-lat].
 %%CITATION = ARXIV:0711.3160;%%"

% #17
\bibitem{b01}
  BELLE Collaboration,
  %``Measurement of B(anti-B0 --> D+ l- anti-nu) and determination of  |V(cb)|''
  Phys.\ Lett.\ B {\bf 526}, 258 (2002).
%  [hep-ex/0111082].
  %%CITATION = HEP-EX/0111082;%%

% #18
\bibitem{HFAG}
Heavy Flavor Averaging Group, [0704.3575][hep-ex],
and online update at http://www.slac.stanford.edu/xorg/hfag/.

% #19
\bibitem{HQET}
N.~Isgur and M.B.~Wise, Phys. Lett. B {\bf232}, 113 (1989); {\bf237}, 527 (1990);
B.~Grinstein, Nucl. Phys. B {\bf339}, 253 (1990);
E.~Eichten and B.~Hill, Phys. Lett. B {\bf234}, 511 (1990);
H.~Georgi, Phys. Lett. B {\bf240}, 447 (1990). 

% #20
\bibitem{n94}
  M.~Neubert,
  %``Heavy Quark Symmetry''
  Phys.\ Rept.\ {\bf245}, 259 (1994).
%  [hep-ph/9306320].
  %%CITATION = HEP-PH/9306320;%%

% #21
\bibitem{bgl}
C.G.~Boyd, B.~Grinstein and R.F.~Lebed,
Phys.\ Rev.\ D {\bf 56}, 6895 (1997).

% #22
\bibitem{cn}
  I.~Caprini, L.~Lellouch and M.~Neubert,
  %``Dispersive Bounds on the Shape of B --> D(*) l nubar Form Factors''
  Nucl.\ Phys.\ B {\bf 530}, 153 (1998).
%  [hep-ph/9712417].
  %%CITATION = HEP-PH/9712417;%%

% #23
% provide eta(1)=0.6+-0.2 used for form factors
\bibitem{lnn}
 Z.~Ligeti, Y.~Nir and M.~Neubert,
 %``The Subleading Isgur-Wise form-factor Xi-3 (v - v-prime) and its
 %implications for the decays anti-B $\to$ D* lepton anti-neutrino'',
 Phys.\ Rev.\ D {\bf 49}, 1302 (1994).
% [hep-ph/9305304].
 %%CITATION = HEP-PH/9305304;%%"

% #24
%explicit formulae for radiative corrections to form factors
\bibitem{n92}
  M.~Neubert,
  %``Short distance expansion of heavy quark currents''
  Phys.\ Rev.\ D {\bf 46}, 2212 (1992).
  %%CITATION = PHRVA,D46,2212;%%

% #25
\bibitem{pdg}
 W.-M. Yao \emph{et al.} [Particle Data Group],
 %``Review of Particle Physics''
 J.\ Phys.\ G {\bf 33}, 1 (2006).
%and 2007 partial update for the 2008 edition.

% #26
\bibitem{rh}
  R.~Hill,
  %``The modern description of semileptonic meson form factors''
  [hep-ph/0606023].
  %%CITATION = HEP-PH/0606023;%%

% #27
\bibitem{FvQuenched}
  G.~M.~de Divitiis, R.~Petronzio and N.~Tantalo,
  %``Quenched lattice calculation of semileptonic heavy-light meson form
  %factors,''
  JHEP {\bf 0710}, 062 (2007).
  %%CITATION = JHEPA,0710,062;%% 

% #28
\bibitem{FvStaggered}
M.~Okamoto \emph{et al.}, Nucl. Phys. Proc. Suppl. {\bf 140}, 461 (2005).

% #29
% subthreshold poles in Hill parametrization
\bibitem{poles}
 E.~Eichten and C.~Quigg,
 %``Mesons with beauty and charm: Spectroscopy'',
 Phys.\ Rev.\ D {\bf 49}, 5845 (1994).
% [hep-ph/9402210].
 %%CITATION = HEP-PH/9402210;%%"

% #30
\bibitem{yuk}
 A.~Buras \emph{et al.},
 %``Delta(M(d,s)), B/(d,s)0 --> mu+ mu- and B --> X/s gamma in
 %supersymmetry at large tan(beta)''
 Nucl.\ Phys.\ B {\bf 659}, 3 (2003).
% [hep-ph/0210145].
 %%CITATION = HEP-PH/0210145;%%

% #31
\bibitem{babarbdt}
  BABAR Collaboration,
Phys.\ Rev.\ Lett. {\bf 100}, 021801 (2008).
%``Observation of the Semileptonic Decays B --> D* tau nubar and 
%Evidence for B --> D tau nubar''
  %%CITATION = ARXIV:0709.1698;%%"

% #32
\bibitem{vub}
% CKM element Vub
CKMfitter, fit inputs for summer 2007, http://www.slac.\linebreak[2] stanford.edu/xorg/ckmfitter/.

% #33
\bibitem{run}
% RunDec and input parameters mc(mc),mb(mb),alphas
 K.~Chetyrkin, J.~K\"uhn and M.~Steinhauser,
 %``RunDec: A Mathematica package for running and decoupling of
 %                 the strong  coupling and quark masses''
 Comput.\ Phys.\ Commun. {\bf 133}, 43 (2000).
% [hep-ph/0004189].
 %%CITATION = HEP-PH/0004189;%%"
Input values:
 J.~K\"uhn, M.~Steinhauser and C.~Sturm,
 %``Heavy quark masses from sum rules in four-loop
 %                 approximation''
 Nucl.\ Phys.\ B {\bf 778}, 192 (2007).
% [hep-ph/0702103].
 %%CITATION = HEP-PH/0702103;%%"

% #34
\bibitem{ccp}
    A.~Bartl, K.~Hohenwarter-Sodek, T.~Kernreiter, O.~Kittel and M.~Terwort,
  %``CP observables with spin-spin correlations in chargino production,''
  [0802.3592][hep-ph],
  %%CITATION = ARXIV:0802.3592;%% 
  and references therein.

% #35
\bibitem{babarbdl}
  BABAR Collaboration,
  %``Measurement of the Branching Fractions of Exclusive anti-B->
  %D/D*/D(*) pi l- anti-nu_l Decays in Events Tagged by a Fully 
  % Reconstructed B Meson''
  [0708.1738][hep-ex].
  %%CITATION = ARXIV:0708.1738;%%"

\end{thebibliography}
\end{document}